\documentclass[aps,prc,showpacs,twocolumn,amssymb,floatfix]{revtex4}
\usepackage{graphicx}

\begin{document}

\title{Transverse momentum spectra of fermions and bosons \\
produced in strong abelian fields}
\author{Vladimir V. Skokov}
\affiliation{Bogoliubov Laboratory of Theoretical Physics, 
Joint Institute for Nuclear Research, \\
141980, Dubna, Russia}
\author{P\'eter L\'evai}
\affiliation{RMKI Research Institute for Particle and Nuclear Physics, \\
P.O. Box 49, Budapest 1525, Hungary}

\date{25 October   2004}

\begin{abstract}
We study the transverse momentum spectra of fermions and
bosons produced in strong, time-dependent abelian field.
The transverse size of the abelian field is finite, similarly to
color strings and ropes. Different time-dependent
field strengths 
are investigated in a kinetic model, and transverse momentum spectra
are calculated for fermions and bosons. These spectra 
display exponential or polynomial behavior at high $p_T$,
depending on the given time dependence.
We compare our spectra to lattice result for a
classical gluon field and obtain surprisingly
good agreement in certain cases.
\end{abstract}


\maketitle

\section{Introduction}

Ultrarelativistic heavy ion collisions at SPS and RHIC energies
($\sqrt{s} = 10 - 200$ AGeV) provided an enormous volume of experimental 
data on particle production~\cite{QM04}. The microscopic
mechanisms of hadron production are not fully understood and
many different descriptions coexist to explain these data.
Transverse momentum spectra of produced hadrons 
have been measured in a wide momentum region 
(e.g. $0 \leq p_T \leq 15$ GeV at RHIC), and can become a decisive test
between different perturbative and non-perturbative models of
hadron formation.

Investigations of $pp$ collisions at high energies led to
the introduction of chromoelectric flux tube ('string') models, where
these tubes are connecting
the quark and diquark constituents
of colliding protons~\cite{FRITIOF,HIJ,RQMD}. 
Strings are good examples of how to convert the kinetic
energy of a collision into field energy. When these flux tubes
become unstable, new hadrons will be produced 
via quark-antiquark and diquark-antidiquark pair production.
Such models can describe experimental data successfully at low $p_T$,
$p_T < 2-3$ GeV. 
At higher $p_T$ perturbative QCD-based models are
working~\cite{FieldFeyn,Wang00,Yi02}.

In heavy ion reactions finite number of nucleons collide and 
the number of produced strings scales with the number of participant
nucleons. Since hadron production at low $p_T$ scales with
participant nucleons in a wide energy range, 
string models could reproduce available data surprisingly well 
in the soft region at SPS energies~\cite{FRITIOF,HIJ,RQMD}.
However, the obtained high density for strings strengthened the idea of
string melting and the formation of color ropes~\cite{Biro}, which fill the 
finite transverse extension partially or completely. 
Following these ideas, 
measured strangeness enhancement was explained successfully
by rope formation~\cite{RQMD}.
This result has indicated the possible formation
of extended color fields at SPS energies.

At RHIC and LHC energies the string density is expected to be so large that
a strong gluon field will be formed in the whole
available transverse volume.
Furthermore, the gluon number will be so high that a classical gluon field
as the expectation value of the quantum field can be considered
and investigated in the reaction volume.
The properties of such non-abelian classical fields and details
of gluon production 
were studied very intensively during the last years, especially
asymptotic solutions~\cite{McL02,IanVen03}.
Fermion production was calculated recently~\cite{Gelis04,Blaiz04}.
Lattice calculations were performed also to describe 
strong classical fields under finite space-time 
conditions~\cite{Krasnitz,Lappi}.

Fermion pair production together with boson pair production
were investigated by kinetic models of particle production from 
strong abelian~\cite{Gatoff87,Kluger91,Gatoff92,Wong95,Eisenberg95,Vinnik02,
Pervu03,Diet03}
and non-abelian~\cite{Prozor03} fields. These calculations concentrated
mostly on bulk properties, the time dependence of energy and particle
number densities.

Our main interest is the transverse momentum distribution of
produced fermions and bosons. Before performing
non-abelian kinetic model calculation, we would like to understand
the role of time dependence, the interplay between production
and annihilation rates in a kinetic approach and the 
influence of finite transverse size on the transverse momentum
distributions. 

In this paper we summarize our results applying a kinetic  model  
with a time-dependent abelian external field characterized by 
finite transverse geometry. We concentrate on transverse
momentum spectra for produced particles.
Section 2 summarizes the field theoretical backgrounds
for boson and fermion production in strong abelian field.
The kinetic equation is explained briefly in Section 3.
In Section 4 the influence of time dependence on fermionic
and bosonic transverse momentum spectra is displayed and 
the obtained results are compared to lattice calculations.
In Section 5 we discuss our results.

\newpage

\section{Bosons and fermions in external field}

Let us consider a massive boson field $\phi$ in an external 
classical abelian vector field, $A^{\mu}$. 
The Lagrangian 
\begin{equation}
{\cal L}_{\phi} = D^*_\mu \phi^* D^\mu \phi - m_+^2 \phi^* \phi 
\end{equation}
leads to the equation of motion
\begin{equation}
(D^\mu D_\mu + m_+^2) \phi = 0 \ , \label{KleinG}
\end{equation}
where $D_\mu = \partial_\mu + ie_+A_\mu$ with bosonic charge $e_+$.
The bosonic mass is labelled by $m_+$.

We will choose a longitudinally dominant vector field in Hamilton gauge with 
the 4-potential
$A^\mu = (0,0,0,A^3),$ 
which is the most appropriate for our investigation in the
ultrarelativistic energy region.

To imitate confinement properties of Yang-Mills fields, the component
$A^3$ is limited in the transverse direction, and a finite 
'flux tube' radius $r_0$ is considered.  The external field is
cylindrically symmetric. 
It vanishes outside the tube, $A^3(t,r>r_0,\varphi,x_3)=0$, and
it is homogeneous inside the flux tube, 
$A^3(t,r \leq r_0,\varphi,x_3)\equiv A^3(t)$~\cite{Gatoff92,Wong95,Eisenberg95}.
 
The Klein-Gordon equation (\ref{KleinG}) reads for the 
boson field $\phi(t,r,\varphi,x_3)$ as
\begin{equation}
\bigg[\partial_0^2 - \nabla_{\perp}^2 - \partial_3^2  -
2ie_+A_3(t) \partial_3+e_+^2A_3^2(t)
+ m_+^2 \bigg] \phi= 0 \ , \label{transkg}
\end{equation}
where the transverse Laplace operator is given by
\begin{eqnarray}
\nabla^2_{\perp}=\frac{\partial^2}{\partial r^2}+
\frac{1}{r}\frac{\partial}{\partial r}
+\frac{1}{r^2}\frac{\partial^2}{\partial \varphi^2}\, .
\end{eqnarray}
For the bosonic field function in
eq.(\ref{transkg}) we are looking for the following
solution:
\begin{eqnarray}
\phi(t,r,\varphi,x_3)=N e^{il\varphi}e^{-ik_3x_3} T(t)R(r)\, .
\label{bosona}
\end{eqnarray}
One equation is obtained for the time-dependent part,
\begin{eqnarray}\label{kgtime}
\ddot T(t) + \left[ m_+^2+ \left(k_3 - e_+ A_3(t) \right)^2 +
\epsilon^2 \right]T(t)=0 \ ,
\end{eqnarray}
and one for the spatial dependence,
\begin{eqnarray}\label{kgspace}
r^2 R^{''} (r) + r R^{'}(r)+\left(\epsilon^2 r^2 - l^2 \right)R(r)=0\, .
\end{eqnarray}
Here $\epsilon$ is the separation constant which will be  fixed later.

Considering flux-tube boundary condition for $A_3(t)$
and the constraint $R(\epsilon r_0)=0$ on the surface of the
flux tube, the solution for the boson field is 
\begin{eqnarray}\label{boson1}
\phi_{nl,k_3}(t,r,\varphi,x_3)=e^{-ik_3x_3} 
\frac{ T_{nl,k_3}(t) \ J_l(\epsilon_{nl} r)}{\sqrt{\pi}
r_0 J^{'}_l(\epsilon_{nl} r_0)} e^{il\varphi}, 
\end{eqnarray}
where $\epsilon_{nl} r_0$ is the $n^{th}$ zero of the Bessel
function $J_l$ and the constant $J^{'}_l(\epsilon_{nl} r_0)$ appears during
the normalization of the field function.

The energy  of the bosonic quasiparticles
reads
\begin{eqnarray}\label{350}
\omega_{+}^2(t,k_T,k_3)&=&m_+^2 + \epsilon _{nl }^{2} + {P_{+}^2}, 
\end{eqnarray}
where $P_{+} = k_3-e_+ A_3(t)$ is the kinetic longitudinal
momenta and $\epsilon _{nl}$ labels the discrete transverse momenta.
In  Section 3  the transverse momenta spectra of the newly produced
bosons will be determined at these discrete transverse momentum values,
$\epsilon _{nl}$.

Massive fermions ($m_-, e_-$) can be described similarly
in the presence of the above external classical abelian field $A^\mu(t)$.
Considering the fermionic Lagrangian 
\begin{equation}
{\cal L}_\psi = {\overline \psi} \left( i \gamma^\mu D_\mu - m_- \right) \psi
\ \ ,
\end{equation}
one obtains the corresponding equation of motion
\begin{equation}
(i \gamma^\mu \partial_\mu -e_-\gamma^3 A_3(t) - m_-) \psi(x) = 0 \ .
\label{ferm0}
\end{equation}
The solution of eq.~(\ref{ferm0}) is wanted in the form of
\begin{eqnarray}
\psi(x) &=& (i \gamma^\mu \partial_\mu -e_-\gamma^3 A_3(t) + m_-) 
\ {\widehat \psi}(x) \ ,
\label{ferm1}
\end{eqnarray}
where an auxiliary field ${\widehat \psi}(x)$ was introduced \cite{grib94}.
In the cylindrically symmetric case, one obtains the following equation for
${\widehat \psi}(t,r,\phi,x_3)$:
\begin{eqnarray}
&&\left[\partial_0^2 - \nabla_{\perp}^2 - \partial_3^2  -
2ie_-A_3(t) \partial_3 +e_-^2A_3^2(t) \right. \nonumber \\
&& {\hspace*{2truecm}} \left. +ie_- \gamma^0 \gamma^3 \partial_0 A_3
+ m_-^2 \right] {\widehat \psi}= 0 \ . \label{transf}
\end{eqnarray}
Now we can look for the solution of eq.(\ref{transf}) in a form similar to
eq.(\ref{bosona}):
\begin{eqnarray}
{\widehat \psi}(t,r,\varphi,x_3)=N e^{il\varphi}e^{-ik_3x_3} {\widehat T}(t)
{\widehat R}(r) \chi_\lambda \, . \label{ferm2}
\end{eqnarray}
Eigenvector of matrix $\gamma_0 \gamma_3$ is labelled by $\chi_\lambda$,
$\lambda=\pm 1$.
During the solution of eq.(\ref{transf}),
the radial equation for ${\widehat R}(r)$
leads to 
eq.(\ref{kgspace}). Thus, one obtains
${\widehat R}(r) \equiv R(r)$ and the same radial solutions with
the same separation constants $\epsilon_{nl}$ as above.
The one-particle energy for the auxiliary field is the following:
\begin{eqnarray}\label{351}
{\widehat \omega}_{-}^2(t,k_T,k_3)&=&
m_-^2 + \epsilon_{nl}^2 +P_{-}^2 \ .
\label{dispf}
\end{eqnarray}
Here $P_{-}(t) = k_3-e_- A_3(t)$ labels longitudinal \mbox{momenta.}
Time-dependent component ${\widehat T}(t)$ 
satisfies the equation
\begin{eqnarray}\label{dirtime}
\ddot {\widehat T}(t) + \left[ {\widehat \omega}^2_{-}(t,k_T,k_3)+
ie_-{\dot A}_3(t) \right] {\widehat T}(t)=0 \ .
\end{eqnarray}

Using eq.(\ref{ferm1}) for the solution of the 
auxiliary field,  one can reconstruct 
the fermion field $\psi(t,r,\varphi,x_3)$. 
The dispersion relation in eq.(\ref{dispf}) remains valid  for these fermions.

The finite radius of the flux tube determines a minimal transverse energy,
$\epsilon_{10} = 0.2404 / r_0$. For $pp$ collisions we obtain 
$\epsilon_{10}(r_0=1\ {\textrm fm}) = 47.4 \ {\textrm MeV}$,
for central $AuAu$ collisions it is 
$\epsilon_{10}(r_0=6.5 \ {\textrm fm}) = 7.3 \ {\textrm MeV}$,
which are very small, but nonzero
values.

\section{The kinetic equation}

The kinetic equation with time-dependent external field can be 
constructed on the basis of the
calculated bosonic and fermionic field functions.
Using a time-dependent Bogoliubov 
transformation approach (see e.g. Refs.~\cite{smol97,grib94}) or
a quasiparticle representation approach~\cite{Pervu03}, one obtains the
following kinetic equation:
\begin{equation}\label{KE}
\frac{{\partial f_\pm (t,k_3)}} {{\partial t}} +
e_\pm E(t)\frac{{\partial f_{\pm}  (t, k_3)}}
{{\partial k_3 }} = {\cal S}_{\pm } (t, k_3).
 \end{equation}
Here  $E(t)=-{\dot A}_3(t)$.
One-particle distribution functions for bosons($+$) and
fermions($-$) are labelled by $f_\pm(t, k_3)$. 
We note that the measurable longitudinal momentum of the produced 
particle has the value of $P_\pm$. 

Equation (\ref{KE}) will be solved at the discrete transverse
momenta $\epsilon_{nl}$, which
depend on the transverse
radius ($r_0$) of the chromoelectric field.
Indeed, eq.~(\ref{KE}) can be solved at continuous transverse momenta
and the solutions at corresponding $\epsilon_{nl}$ will be selected.

The source term ${\cal S}_{\pm}(t, k_3)$ contains information about the
time evolution of the system and is determined as
\begin{eqnarray} \label{KEsource}
{\cal S}_{\pm }(t,k_3) = \frac{1}{2} W_{\pm} (t,k_3) \int\limits_{- \infty }^t
{dt'}\,  \,  {\overline W}_{\pm }(t,t';k_3)\cdot
&& \nonumber\\
\left\{ \left[1 \pm 2f_\pm(t',k_3) 
\right] 
\cos \left[ {2 \int\limits_{t' }^t {d \tau}
\ {\overline \omega}_{\pm}(t, \tau; k_3)}\right] \right\}, \  &&
\end{eqnarray}
where
\begin{eqnarray}
{{{\overline \omega}_{\pm }^2 (t,t'; k_3)}} &=& 
m_{\pm}^2 + \epsilon_{nl}^2 + {\overline P}_\pm^2 (t,t')
\, , \ \ \label{opm} \\
{\overline P}_\pm (t,t')& =& 
k_3  + e_\pm \int\limits_{t' }^t {d\tau E(\tau )}\, , 
\label{ppm} \\
{\overline W}_{\pm } (t,t'; k_3 ) &=& 
\frac{{e_\pm E(t){\overline P}_\pm (t,t' )}}
{{{\overline \omega}_{\pm }^2 (t,t'; k_3)}} 
\left[ {\frac{\sqrt{m_\pm^2+\epsilon_{nl}^2} }
{{{\overline P}_\pm (t,t')}}} \right]^{2s_\pm}\, . \ \ \label{wpm}  
 \end{eqnarray}
 
Overlined notation indicates appropriate asymptotic behavior,
namely $W_{\pm}(t,k_3) \equiv {\overline W}_{\pm } 
(t,t'\rightarrow -\infty; k_3 )$.
Similar asymptotics are valid for $P_{\pm}(t)$ and $\omega_\pm(t)$.

The difference in boson and fermion production appears through the
spin factor in eq.~(\ref{wpm}), namely
$s_+=0$ for bosons and $s_-=1/2$ for fermions.

Now we introduce new auxiliary functions, $v_\pm$ and $u_\pm$,  and
rewrite the integro-differential eq.(\ref{KE}) in a more convenient form
containing full derivatives:
\begin{eqnarray}
{\dot f}_\pm &=& \frac{1}{2} W_\pm v_\pm \ ,  \label{KE1}\\
{\dot v}_\pm &=& W_\pm (1 \pm 2 f_\pm) - 2 \omega_\pm u_\pm \ , \label{KE2} \\
{\dot u}_\pm &=& 2 \omega_\pm v_\pm \ . \label{KE3}
\end{eqnarray}
This system of differential equations is solved for all sets of 
$\{ \epsilon_{nl}, k_3  \}$,
which correspond to the different transverse and longitudinal momenta.

The functions $f_\pm, v_\pm, u_\pm$ start with zero initial value,
describing a system without any bosons or fermions at the beginning. 
Assuming different time dependences for the 
external field, $E(t)$, we obtain different particle productions.
We would like to note that produced charged particles
 generate an inner field which interferes with the external 
one~\cite{Pervu03}. This effect is neglected here, as well as
a possible collision term, ${\cal C}_\pm(t,k_3)$, on the right-hand
side of eq.~(\ref{KE})~\cite{skokov02}.
A given external field can
mimic a certain self-consistent solution
as expansion, collision or back reaction influenced 
time dependence of the field.
\begin{figure}[b]
\centerline{%
\rotatebox{0}{\includegraphics[height=7.0truecm]%
   {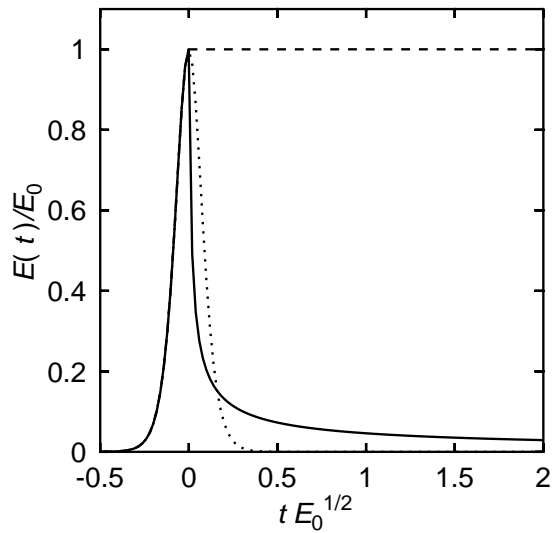}}
}
\caption{
The time dependence of
external field $E(t)$ in three physical scenarios:
a) pulse ({\it dotted line});
b) constant field, $E_0$ ({\it dashed line});
c) scaled decrease ({\it solid  line}).
}
\label{etime}
\end{figure}

\section{Numerical results}

In heavy ion collisions, one can assume three different types of
time dependence
for the chromoelectric field to be formed: 
a) pulse-like field develops with a fast increase, 
which is followed by a fast fall in the field strength;
b) formation of a constant field
($E_0$) is maintained after the fast increase in the initial time period;
c) scaled decrease of the field strength appears, 
which is caused by
particle production and/or transverse expansion, and the decrease
is elongated in time much further than the pulse-like assumption.

Figure ~\ref{etime} displays three sets for the
time dependence of the external field to investigate the
above three different physical scenarios numerically: 
\begin{eqnarray}
E_{pulse}(t) &= E_0 \cdot \left[ 1- \textrm{tanh} (t/\delta) \right] 
\hspace*{0.6truecm}   & \label{Ep} \\
E_{const}(t) &= \left\{ \begin{array}{ll}
     E_{pulse}(t) & {\textrm at} \ \ {t < 0} \\  
     E_0          & {\textrm at} \ \ {t \geq 0} 
                \end{array} \right. \label{Ec} \\
E_{scaled}(t) &= \left\{ \begin{array}{ll}
     E_{pulse}(t) & {\textrm at} \ \ {t < 0} \\ 
     \frac{E_0}{(1+t/t_0)^{\kappa}}          & {\textrm at} \ \ {t \geq 0} 
                \end{array} \right. \label{Es}
\end{eqnarray}

Here we choose $\delta=0.1 / \, E_0^{1/2}$, which corresponds to RHIC energies,
and $\kappa=2/3$ to indicate a  longitudinally scaled Bjorken expansion
with $t_0 = 0.01 / \, E_0^{1/2}$. 
In fact, the whole time dependence
is scaled by ${E_0}^{1/2}$.

\begin{figure}[t]
\centerline{%
\rotatebox{0}{\includegraphics[height=7.0truecm]%
   {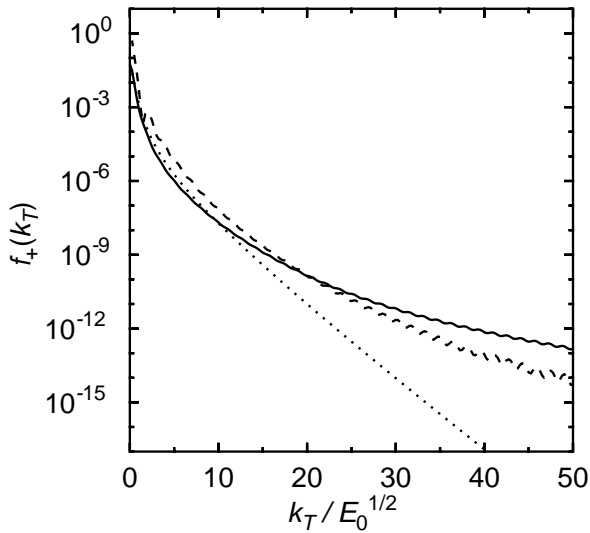}}
}
\caption{
The transverse momentum spectra for bosons in the three physical
scenarios (see Fig.~\ref{etime} and text for explanation).
}
\label{bostime}
\end{figure}

\begin{figure}[b]
\centerline{%
\rotatebox{0}{\includegraphics[height=7.0truecm]%
   {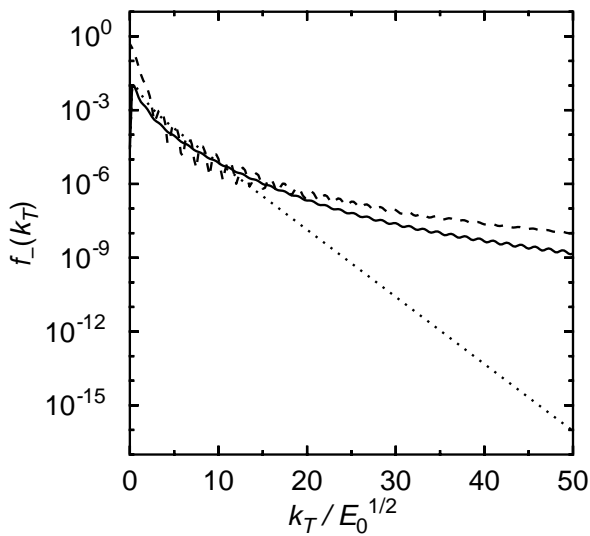}}
}
\caption{
The transverse momentum spectra for fermions in the three physical
scenarios (see Fig.~\ref{etime} and text for explanation).
}
\label{fertime}
\end{figure}

Figures \ref{bostime} and \ref{fertime}
display the obtained transverse momentum spectra in midrapidity
at  time $t=2/E_0^{1/2}$ for bosons
and fermions.
Pulse-type time dependence leads to exponential spectra ({\it dotted lines}),
$f_\pm \propto \exp (k_T/T_\pm)$, where
the slope value is $T_+ = 1.6 \cdot  E_0^{1/2}$ for bosons and
$T_- = 1.45 \cdot  E_0^{1/2}$ for fermions. 
In the other two cases,  we obtain non-exponential spectra, which are
generated by the long-lived electric field. 
The transverse spectra from constant 
({\it dashed lines}) and scaled ({\it solid lines}) fields are close,
because the production and annihilation rates 
balance each other.  Slight differences appear because
of the fast fall of the scaled field  immediately after $t = 0$.
We note here, 
that indeed we have discrete transverse momenta, $\epsilon_{nl}$,
but with such small steps, which allow us to display continuous
transverse momentum spectra. 

Comparing the boson and fermion production in midrapidity, 
one can recognize the large difference in the production rates 
and see a clear fermion dominance at all transverse momenta. 
The reason of this dominance can be understood from
Fig. 4, which displays the longitudinal particle spectra, $f_\pm(P_\pm)$,
at fixed transverse momenta, $k_T/E_0^{1/2} = 1$, and time, $t= 2/ E_0^{1/2}$. 
Fermion spectra (thick line) have a reasonable value around zero 
longitudinal momenta (which means midrapidity), 
but boson spectra (thin line) display a 'dip'-like structure here.
The presence of this 'dip' is related to
the structure of the transition rate ${\overline W}(t,t';k_3)$ 
in eq.~(\ref{wpm}) and
the longitudinal dominance of the electric field defined via $A_3(t)$: 
at zero longitudinal momenta ($P_\pm$) 
there is no boson production, but fermion production
is characterized by a well defined finite value.

\begin{figure}[t]
\centerline{%
\rotatebox{0}{\includegraphics[height=7.0truecm]%
   {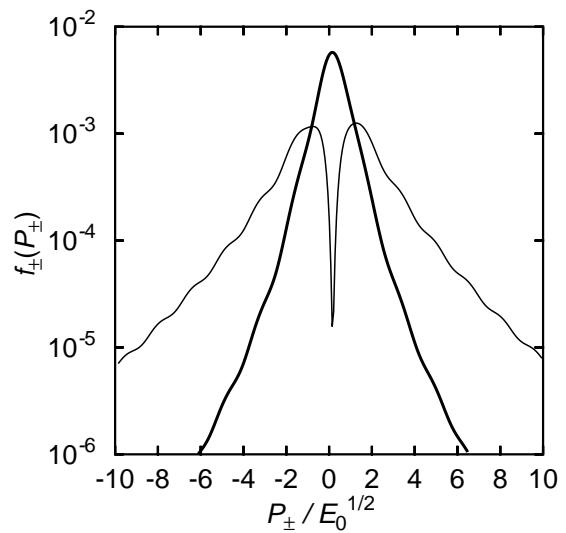}}
}
\caption{
Longitudinal momentum spectra for fermions (thick line)
and bosons (thin line) at $k_T/E_0^{1/2} = 1$
and $t= 2/ E_0^{1/2}$.
}
\label{long}
\end{figure}

At extremely high-energy heavy ion collisions we expect this longitudinal
dominance of the chromoelectric field.
If transverse electric field components become comparable to the longitudinal 
one at lower energies, then the boson yield will be much larger. 
The presence of scattering
term, ${\cal C}_\pm$, can increase the yield in midrapidity from the
neighbouring rapidity cells, where the yield is large, even
if a longitudinal field component, $A^3$, exists alone.

\begin{figure}[t]
\centerline{%
\rotatebox{0}{\includegraphics[height=7.0truecm]%
   {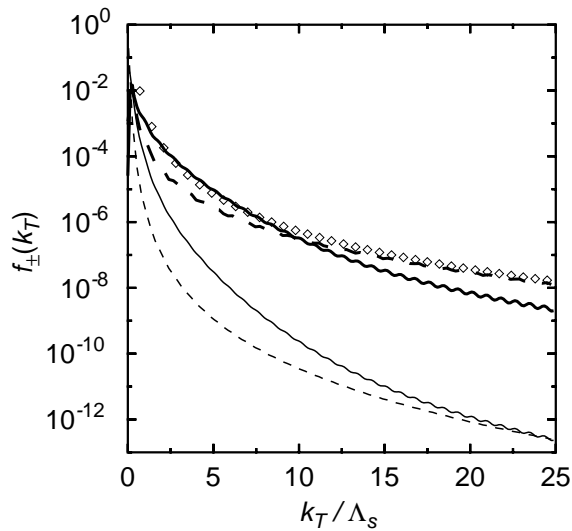}}
}
\caption{
Transverse momentum spectra for fermions ({\it thick lines}) 
and bosons ({\it thin lines}) from our calculation with scaled
time evolution, $E_{scaled}(t)$, 
 and gluon production ({\it diamonds}) from SU(3) 
 lattice-QCD calculation~\cite{Krasnitz}.
}
\label{kras}
\end{figure}

In Refs.~\cite{Krasnitz} transverse momentum spectra for gluons
were determined in SU(3) classical field theory. Comparison between
this non-abelian and our abelian calculation can be made, if we fix 
both field strength at the same value. 
The energy density of the classical gluon field is given in 
Refs.~\cite{Krasnitz} as
$\varepsilon_{SU3} = 0.17 \Lambda_s^4 / g^2$, where 
$g=2\  (\alpha_s = 0.33)$. In our calculation, the initial energy density is
related to the field strength $E_0(t=0)$:
$\varepsilon = E_0^2 / 2$. Assuming the same 'initial' energy density in the
abelian and the non-abelian case, we obtain $E_0^{1/2}/\Lambda_s = 0.54$.
Now we can rescale our transverse momentum spectra  from
Figures {\ref{bostime}} and {\ref{fertime}} - we choose the results
with scaled time-dependent field, $E_{scaled}$ (solid lines).
In Figure \ref{kras} our results at RHIC energies 
for fermion ({\it thick solid line})
and boson ({\it thin solid line}) are displayed together with  non-abelian
result for classical gluons~\cite{Krasnitz} ({\it open diamonds}).
Our fermion spectra and the lattice result on 
classical gluons overlap
in the momentum region $2 < k_T / \Lambda_s < 10$.
Possible parametrization of the two spectra are very close to each other
and display the $\log(k_T/\Lambda_s) \cdot (k_T/\Lambda_s)^{-4}$
behavior known from Ref.~\cite{Krasnitz} and from 
perturbative QCD~\cite{Gyul97}. 
An acceptable slight modification of the time evolution
in eq.~(\ref{Es})  could lead to a full overlap, even at high-$k_T$.
This can be demonstrated at $k_T / \Lambda_s > 15$
with  $\delta=0.01 / \, E_0^{1/2}$ for fermion production
({\it thick dashed line}).

\section{Discussion}

In this paper we investigated fermion and boson production from
a strong abelian field in a kinetic model. Assuming ultrarelativistic 
limit and a longitudinally dominant external field,
boson production remains relatively weak in the mid-rapidity region.
However, fermion production is not effected by this setup and a
fermion dominance appears in our kinetic model.
Introducing different parametrizations for the time evolution
of the external field, we obtain different transverse momentum
spectra for fermions and bosons. Only time evolution determines
if these spectra are exponential or polynomial.
Investigating fermion transverse momentum spectra, our results
with abelian external field overlap with the spectra obtained
in lattice-QCD calculations for classical gluons at the same
energy density.

\section*{Acknowledgments}

We thank V. Toneev and G. Fai for stimulating discussions.
This work was supported in part by  Hungarian grant OTKA-T043455,
MTA-JINR Grant, and RFBR grant No. 03-02-16877.

\end{document}